\documentclass{llncs}

\usepackage{color}
\usepackage{cite}
\usepackage{amsmath}
\usepackage{amsfonts}
\usepackage{graphicx} 
\usepackage{bm}
\usepackage{epstopdf}
\usepackage{xcolor,colortbl}
\usepackage{subfigure}
\definecolor{Gray}{gray}{0.85}

\begin{document}

\title{Design of acoustic metamaterials through nonlinear programming}

\author{Andrea Bacigalupo\inst{1}, Giorgio Gnecco\inst{1}, Marco Lepidi\inst{2}, Luigi Gambarotta\inst{2}}
\institute{IMT School for Advanced Studies, Lucca, Italy \and DICCA, University of Genoa, Italy \\
\email{\{andrea.bacigalupo,giorgio.gnecco\}@imtlucca.it, \{marco.lepidi,luigi.gambarotta\}@unige.it}}

\maketitle

\begin{abstract}

The dispersive wave propagation in a periodic metamaterial with tetrachiral topology and inertial local resonators is investigated. The Floquet-Bloch spectrum of the metamaterial is compared with that of the tetrachiral beam lattice material without resonators. The resonators can be designed to open and shift frequency band gaps, that is, spectrum intervals in which harmonic waves do not propagate. Therefore, an optimal passive control of the frequency band structure can be pursued in the metamaterial. To this aim, a suitable constrained nonlinear optimization problem on a compact set of admissible geometrical and mechanical parameters is stated. According to functional requirements, the particular set of parameters which determines the largest low-frequency band gap between a pair of consecutive branches of the Floquet-Bloch spectrum is obtained. The optimization problem is successfully solved by means of a version of the method of moving asymptotes, combined with a quasi-Monte Carlo multi-start technique. 
\end{abstract}


\section{Introduction}\label{sec:Introduction}

An increasing interest has been recently attracted by the analysis of the transmission and dispersion properties of the elastic waves propagating across periodic materials \cite{Phani2006,Tee2010,Spadoni2009,BacigalupoLepidi2015,Bacigalupo2014a,BacigalupoDeBellis2015,BacigalupoLepidi2016}. In particular, several studies have been developed to parametrically assess the dispersion curves characterizing the wave frequency spectrum and, therefrom, the amplitudes and boundaries of frequency band gaps lying between pairs of consecutive non-intersecting dispersion curves.

In this background, a promising improvement with respect to conventional beam lattice materials, realized by a regular pattern of stiff disks/rings connected by flexible ligaments, consists in converting them into {\em inertial metamaterials}. To this aim, intra-ring inertial resonators, elastically coupled to the microstructure of the beam lattice material are introduced \cite{Liu2011,Tan2012,Bigoni2013,BacigalupoGambarotta2015}. If properly optimized, the geometrical and mechanical parameters of the metamaterial may allow the adjustment and enhancement of acoustic properties. For instance, challenging perspectives arise in the tailor-made design of the frequency spectrum for specific purposes, such as opening, enlarging, closing or shifting band gaps in target frequency ranges. Once completed, this achievement potentially allow the realization of a novel class of fully customizable mechanical filters. 

Among the others, an efficient approach to the metamaterial design can be based on the formulation of a suited constrained nonlinear optimization problem. In the paper, focus is made on the filtering properties of the tetrachiral periodic material and the associated metamaterial, by seeking for optimal combinations of purely mechanical and geometrical parameters. The relative maximum amplitude of band gaps between different pairs of dispersion curves is sought. This approach strengthens the results already achieved in \cite{Bacigalupoetal2016}, where similar optimization strategies were applied to the passive control of hexachiral beam lattice metamaterials, while the optimization was restricted to the lowest band gap of the Floquet-Bloch spectrum \cite{Brillouin1953} (namely, that lying between the second acoustic branch, and the first optical branch).  The resulting optimization problems are solved numerically by combining one version of the method of moving asymptotes \cite{Svanberg1987} with a quasi-Monte Carlo initialization technique.

The paper is organized as follows. Section \ref{sec:2} describes the physical-mathemati\-cal model of the metamaterial. Section \ref{sec:3} states the relative band gap optimization problem, describes the solution approach adopted, and reports the related numerical results. Finally, Section \ref{sec:4} presents some conclusions. Mechanical details about the physical-mathematical model are reported in the Appendix.

\begin{figure}[b!]

\vspace{-5 pt}
\includegraphics[scale=0.82]{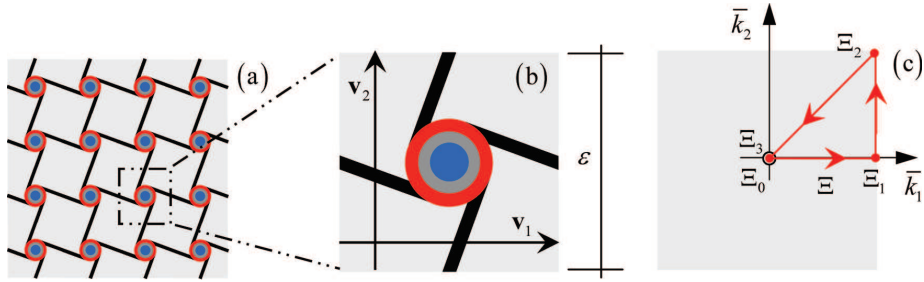}
\caption{Tetrachiral cellular material equipped with resonators: (a) pattern; (b) details of the single cell for the case of tangent ligaments, corresponding to $\beta=\arcsin\left( 2\frac{R}{\varepsilon}\right)$.}
 \label{MODFig01}
\end{figure}

\section{Physical-mathematical model}\label{sec:2}

A planar cellular metamaterial, composed of a periodic tesselation of square cells along two orthogonal periodicity vectors ${\bf v}_1$ and ${\bf v}_2$, is considered. In the absence of an embedding soft matrix, the internal microstructure of each cell, as well as the elastic coupling between adjacent cells, are determined by a periodic pattern of central rings connected to each other by four elastic ligaments, spatially organised according to a tetrachiral geometric topology (see Fig. \ref{MODFig01}a).

Focusing on the planar microstructure with unit thickness of the generic cell (Fig. 1b), the central massive and highly-stiff ring is modelled as a rigid body (in red), characterized by mean radius $R$  and width $w_{an}$. The light and highly-flexible ligaments (in black) are modelled as massless, linear, extensible, unshearable beams, characterized by natural length $L$ (between the ring-beam joints), transversal width $w$, and inclination $\beta$  (with respect to the $\varepsilon$-long line connecting the centres of adjacent rings). By virtue of the periodic symmetry, the cell boundary crosses all the ligaments at midspan, halving their natural length. A heavy internal circular inclusion with external radius $r$ (blue circle in Fig.\ref{MODFig01}b), is located inside the ring through a soft elastic annulus (in grey). This inclusion, modelled as a rigid disk, plays the role of a low-frequency resonator. The beam material is supposed linearly elastic, with Young's modulus $E_s$, and uniform mass density, assumed as negligible with respect to the density $\rho_{s\_an}$ of the highly-stiff ring. Hence, the whole mass of the lattice is assumed to be assigned to the highly-stiff rings. The soft coating inside the resonator is considered a homogeneous, linearly elastic and isotropic solid, with Young's modulus $E_r$  and Poisson's ratio $\nu_r$. It is worth noting that the ligament natural length $L$ is a $(\varepsilon, \beta, R)$-dependent parameter, obeying to the geometric relation
\begin{equation}
L=\varepsilon \left( \cos \beta -\sqrt{\left( {2 R}/{\varepsilon} \right)^2 - \left(\sin\beta\right)^2}\;\right)
\label{MODSeq010}
\end{equation}
Specializing the approach proposed in \cite{Bacigalupoetal2016} to deal with the case-study under investigation, in which the tetrachiral cell topology is featured by two periodicity vectors, the generalized eigenproblem governing the free propagation of harmonic waves (with frequency $\bar{\omega}_h$ and wavevector $\mathbf{\bar{k}}$) in the  metamaterial reads 
\begin{equation}
\left( \mathbf{\bar{K}}\left( \boldsymbol{\bar{\mu }},\mathbf{\bar{k}} \right)-\bar{\omega }_h^2\left( \boldsymbol{\bar{\mu }},\mathbf{\bar{k}} \right) \mathbf{\bar{M}}\left( {\mathbf{\boldsymbol{\bar{\mu }}}} \right) \right)\boldsymbol{\bar{\psi }}\left( \boldsymbol{\bar{\mu }},\mathbf{\bar{k}} \right)=\mathbf{0}
\label{MODSeq020}
\end{equation}
where the dimensionless six-by-six matrices $\mathbf{\bar{K}}\left( \boldsymbol{\bar{\mu }},\mathbf{\bar{k}} \right)$ and $\mathbf{\bar{M}}\left( {\boldsymbol{\bar{\mu }}} \right)$ are Hermitian and diagonal, respectively, and explicitly depend on the minimal dimensionless vector $\boldsymbol{\bar{\mu }}$ of independent geometrical and mechanical parameters 
\begin{equation}
\boldsymbol{\bar{\mu }}=\left( \frac{w}{\varepsilon},\frac{w_{an}}{w},\frac{R}{\varepsilon},\beta,
\frac{r}{\varepsilon},\frac{E_r}{E_s},\nu_r,
\frac{\rho_r}{\rho_{s\_an}}\right) \in \mathbb{R}^{8}
\end{equation}
with components $\bar{\mu}_l$, $l=1,\ldots,8$.

Fixed a certain dimensionless wave vector $\mathbf{\bar{k}} \in \mathbb{R}^2$, the eigenproblem solution is composed by six real-valued eigenvalues $\bar{\omega }_h^2\left( \boldsymbol{\bar{\mu }},\mathbf{\bar{k}} \right)$ ($h=1,\ldots,6$), and the corresponding complex-valued eigenvectors $\boldsymbol{\bar{\psi }}_h\left( \boldsymbol{\bar{\mu }},\mathbf{\bar{k}} \right) \in \mathbb{C}^6$. Here, $\bar{\omega }_h \left( \boldsymbol{\bar{\mu }},\mathbf{\bar{k}} \right)$ is the $h$-th normalized angular frequency, which is related to the unnormalized angular frequency $\omega$ through $\bar{\omega}=\left(\omega \varepsilon\right)/\left({E_s/\rho_{s\_an}}\right){}^{1/2}$. The nonlinear dispersion relations $\bar{\omega }_h\left( \boldsymbol{\bar{\mu }},\mathbf{\bar{k}} \right)$ are the six roots of the nonlinear equation imposing the singularity condition on the matrix governing the linear eigenproblem \eqref{MODSeq020}.

Introducing a suited partition $\boldsymbol{\bar{\psi }}=(\boldsymbol{\bar{\psi }}^s,\boldsymbol{\bar{\psi }}^r)$ of the model degrees-of-freedom, the matrices $\mathbf{\bar{K}}\left( \boldsymbol{\bar{\mu }},\mathbf{\bar{k}} \right)$ and $\mathbf{\bar{M}}\left( {\boldsymbol{\bar{\mu }}} \right)$ have the form
\begin{equation}
\mathbf{\bar{K}}\left(\boldsymbol{\bar{\mu }},\mathbf{\bar{k}} \right)=\left[\;\begin{matrix}
{\mathbf{\bar{K}}}^{s}\left( \boldsymbol{\bar{\mu }},\mathbf{\bar{k}} \right) & \mathbf{\bar{K}}^{sr}\left( \boldsymbol{\bar{\mu }},\mathbf{\bar{k}} \right)  \\[2pt]
\mathbf{\bar{K}}^{rs}\left( \boldsymbol{\bar{\mu }},\mathbf{\bar{k}} \right) & \mathbf{\bar{K}}^{r}\left( \boldsymbol{\bar{\mu }},\mathbf{\bar{k}} \right)  \\
\end{matrix}\;\right],\quad \mathbf{\bar{M}}\left( {\boldsymbol{\bar{\mu }}} \right)=\left[\;\begin{matrix}
 {{{\mathbf{\bar{M}}}}^{s}}\left( {\boldsymbol{\bar{\mu }}} \right) & \mathbf{O}  \\[2pt]
 \mathbf{O} & {{{\mathbf{\bar{M}}}}^{r}}\left( {\boldsymbol{\bar{\mu }}} \right)  \\
\end{matrix}\;\right]
\end{equation}
where the entries of the three-by-three submatrices are reported in the Appendix. The submatrices $\mathbf{\bar{K}}^{sr}\left( \boldsymbol{\bar{\mu }},\mathbf{\bar{k}} \right)$ and $\mathbf{\bar{K}}^{rs}\left( \boldsymbol{\bar{\mu }},\mathbf{\bar{k}} \right)$ describe the interaction between the resonator and the rest of the microstructure.
In the absence of the resonator (i.e., when $r/\varepsilon=0$), the parameter vector reduces to 
\begin{equation}
\boldsymbol{\bar{\mu }}^s=\left( \frac{w}{\varepsilon},\frac{w_{an}}{w},\frac{R}{\varepsilon},\beta \right) \in \mathbb{R}^{4}
\end{equation}
and the generalized eigenvalue problem reduces to
\begin{equation}
\left( {\mathbf{\bar{K}}}^{s}\left( \boldsymbol{\bar{\mu }}^s,\mathbf{\bar{k}} \right) -\bar{\omega }_h^2\left( \boldsymbol{\bar{\mu }}^s,\mathbf{\bar{k}} \right) \mathbf{\bar{M}}^s \left( {\boldsymbol{\bar{\mu }}^s} \right) \right)\boldsymbol{\bar{\psi }}_h^s\left( \boldsymbol{\bar{\mu }}^s,\mathbf{\bar{k}} \right)=\mathbf{0}
\end{equation}
where $h=1,\ldots,3$, and $\boldsymbol{\bar{\psi }}_h^s\left( \boldsymbol{\bar{\mu }}^s,\mathbf{\bar{k}} \right) \in \mathbb{C}^3$.

For any fixed choice of the parameter vector $\boldsymbol{\bar{\mu }}$, the $h$-th dimensionless angular frequency locus  along the closed boundary $\partial B$ of the Brillouin irreducible zone $B$ \cite{Brillouin1953}, spanned anticlockwise by the dimensionless curvilinear coordinate ${\rm \Xi}$ (Fig. \ref{MODFig01}c), is the $h$-th dispersion curve of the Floquet-Bloch spectrum. In particular, the $B$-vertices are $\mathbf{\bar{k}}_0=(0,0)$, $\mathbf{\bar{k}}_1=\left(0,\pi \right)$, $\mathbf{\bar{k}}_2=\left(\pi,\pi \right)$, and $\mathbf{\bar{k}}_3=\mathbf{\bar{k}}_0$. The segments $\partial B_1$ and $\partial B_3$ of the boundary $\partial B$ join, respectively, $\mathbf{\bar{k}}_0$ and $\mathbf{\bar{k}}_1$ (i.e., ${\rm \Xi} \in \left[{\rm \Xi}_0, {\rm \Xi}_1 \right]$ (where ${\rm \Xi}_0=0$ and ${\rm \Xi}_1=\pi$), and $\mathbf{\bar{k}}_2$ and $\mathbf{\bar{k}}_3$ (i.e., ${\rm \Xi} \in \left[{\rm \Xi}_2, {\rm \Xi}_3 \right]$ (where ${\rm \Xi}_2=2 \pi$ and ${\rm \Xi}_3=2 \pi + \sqrt{2}\pi$). For $k=h+1$, the relative amplitude of the full band gap between the $h$-th and $k$-th dispersion curves is
\begin{equation}
\label{eq:relativebandgap}
\Delta {{\bar{\omega }}_{hk,\partial B, \text{rel}}}\left( {\boldsymbol{\bar{\mu }}} \right)=\frac{{{\max }_{\mathbf{\bar{k}}\in \partial {B}}}{{{\bar{\omega }}}_{h}}(\boldsymbol{\bar{\mu }},\mathbf{\bar{k}})-{{\min }_{\mathbf{\bar{k}}\in \partial {B}}}{{{\bar{\omega }}}_{k}}(\boldsymbol{\bar{\mu }},\mathbf{\bar{k}})}{\frac{1}{2}\left[ {{\max }_{\mathbf{\bar{k}}\in \partial {B}}}{{{\bar{\omega }}}_{h}}(\boldsymbol{\bar{\mu }},\mathbf{\bar{k}})+{{\min }_{\mathbf{\bar{k}}\in \partial {B}}}{{{\bar{\omega }}}_{k}}(\boldsymbol{\bar{\mu }},\mathbf{\bar{k}}) \right]}\,.\end{equation}
Partial relative band gaps $\Delta {{\bar{\omega }}_{h, \partial B_1, \text{rel}}}$ are obtained by replacing $\partial B$ with $\partial B_1$ in \eqref{eq:relativebandgap}, and are associated to waves characterized by $\bar{k}_2=0$ and variable $\bar{k}_1$. 
The relative amplitude of the fluctuation of the $h$-th dispersion curve is defined
\begin{equation}
\label{eq:relativeamplitude}
\Delta_A {{\bar{\omega }}_{h, \partial B, \text{rel}}}\left( \boldsymbol{\bar{\mu }} \right)=\frac{{{\max }_{\mathbf{\bar{k}}\in \partial {B}}}{{{\bar{\omega }}}_{h}}(\boldsymbol{\bar{\mu }},\mathbf{\bar{k}})-{{\min }_{\mathbf{\bar{k}}\in \partial {B}}}{{{\bar{\omega }}}_{h}}(\boldsymbol{\bar{\mu }},\mathbf{\bar{k}})}{\frac{1}{2}\left[ {{\max }_{\mathbf{\bar{k}}\in \partial {B}}}{{{\bar{\omega }}}_{h}}(\boldsymbol{\bar{\mu }},\mathbf{\bar{k}})+{{\min }_{\mathbf{\bar{k}}\in \partial {B}}}{{{\bar{\omega }}}_{h}}(\boldsymbol{\bar{\mu }},\mathbf{\bar{k}}) \right]}\,.\end{equation}
To preserve the structural meaning of the solution with proper bounds fixed a priori, the following geometrical constraints on the parameters are introduced
\begin{equation}\label{eq:constraint1}
\frac{1}{10} \frac{R}{\varepsilon} \leq \frac{w_{an}}{w} \frac{w}{\varepsilon} \leq \frac{R}{\varepsilon}\,,
\end{equation}
\begin{equation}\label{eq:constraint2}
\beta \leq \arcsin \left( 2 \frac{R}{\varepsilon} \right)\,,
\end{equation}
\begin{equation}\label{eq:constraint3}
\frac{w}{\varepsilon} \leq \frac{2}{3} \left(\frac{R}{\varepsilon} + \frac{1}{2} \frac{w_{an}}{w} \frac{w}{\varepsilon} \right)\,,
\end{equation}
\begin{equation}\label{eq:constraint4}
\frac{1}{5} \left(\frac{R}{\varepsilon} - \frac{1}{2} \frac{w_{an}}{w} \frac{w}{\varepsilon} \right) \leq \frac{r}{\varepsilon} \leq \frac{9}{10}\left(\frac{R}{\varepsilon} - \frac{1}{2} \frac{w_{an}}{w} \frac{w}{\varepsilon} \right)\,.
\end{equation}
and the related admissible ranges of the parameters are summarized in Table 1. In the absence of the resonator, the definitions \eqref{eq:relativebandgap}, \eqref{eq:relativeamplitude} hold with $\boldsymbol{\bar{\mu }}$ replaced by $\boldsymbol{\bar{\mu }}^s$ and the constraint (\ref{eq:constraint4}) is absent.

\begin{table}
\begin{center}
\setlength{\tabcolsep}{10.5pt}
\renewcommand{\arraystretch}{1.5}
\begin{tabular}{c c c c c c c c c}
\hline
\rowcolor{Gray}
$\rule{-1pt}{12pt}\bar{\mu}_l$ & $\frac{w}{\varepsilon}$ & $\frac{w_{an}}{w}$ & $\frac{R}{\varepsilon}$ & $\beta$ & $\frac{r}{\varepsilon}$ & $\frac{E_r}{E_s}$ & $\nu_r$ & $\frac{\rho_r}{\rho_{s\_an}}\rule[-8pt]{-1pt}{12pt}$ \\
\hline
$\rule{-1pt}{12pt}\bar{\mu}_{l, {\rm min}}$ & $\frac{3}{50}$ & $\frac{1}{20}$ & $\frac{1}{10}$ & $0$ & $\frac{1}{100}$ & $\frac{1}{10}$ & $\frac{1}{5}$ & $\frac{1}{2}\rule[-8pt]{-1pt}{12pt}$ \\
\hline
$\rule{-1pt}{12pt}\bar{\mu}_{l, {\rm max}}$ & $\frac{1}{5}$ & $\frac{10}{3}$ & $\frac{1}{5}$ & $\arcsin \left( \frac{2}{5} \right)$ & $\frac{9}{50}$ & $1$ & $\frac{2}{5}$ & $2\rule[-8pt]{-1pt}{12pt}$ \\
\hline
\end{tabular}
\end{center}
\caption{Lower and upper bounds on the geometrical and mechanical parameters.}
\vspace{-10 pt}
\end{table}

\section{Optimization problems}\label{sec:3}

Some optimization problems, imposed on the Floquet-Bloch spectrum $\bar{\omega }_h(\boldsymbol{\bar{\mu }},\mathbf{\bar{k}})$ of the material/metamaterial, are considered. They are formulated as constrained nonlinear optimization problems, solved by using a version of the method of moving asymptotes \cite{Svanberg1987}, combined with a quasi-Monte Carlo multi-start technique.  Loosely speaking, the solution method consists in tackling a sequence of concave-maximization subproblems, locally approximating the original nonlinear optimization problem\footnote[1]{The {\em moving asymptotes} are asymptotes of functions (changing when moving from one optimization subproblem to the successive one), which are used to approximate the original objective and constraint functions.} (a different approximation at each sequence iteration), where\-as the quasi-Monte Carlo multi-start technique increases the probability of finding a global maximum point through a set of quasi-random initializations of the sequence. More details about the combined method are reported in \cite{Bacigalupoetal2016}.

\subsection{Band gap between the second and third dispersion curves}

\begin{figure}[t]
\subfigure[]{\includegraphics[scale=0.33]{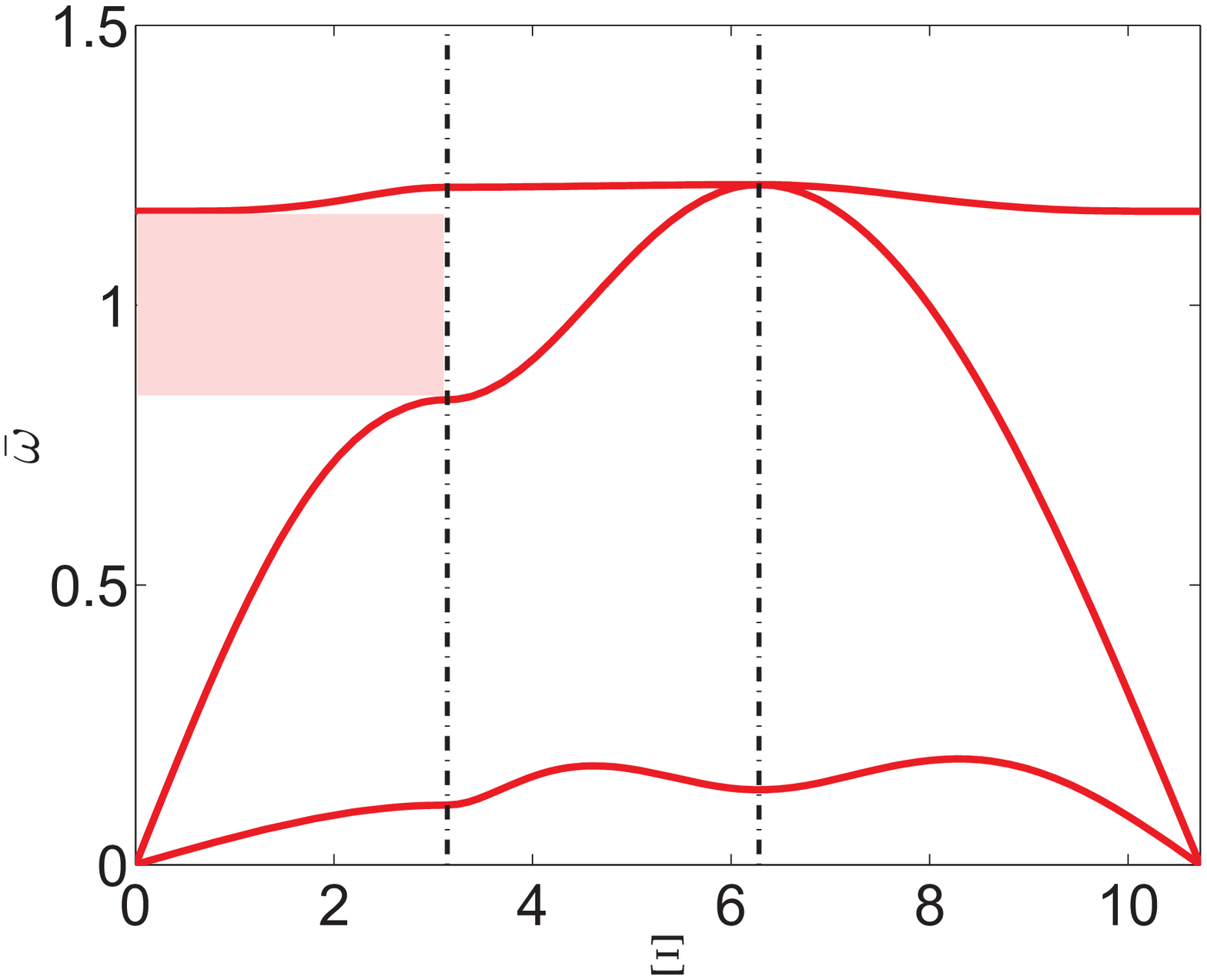}}
\hspace{0.5cm}
\subfigure[]{\includegraphics[scale=0.33]{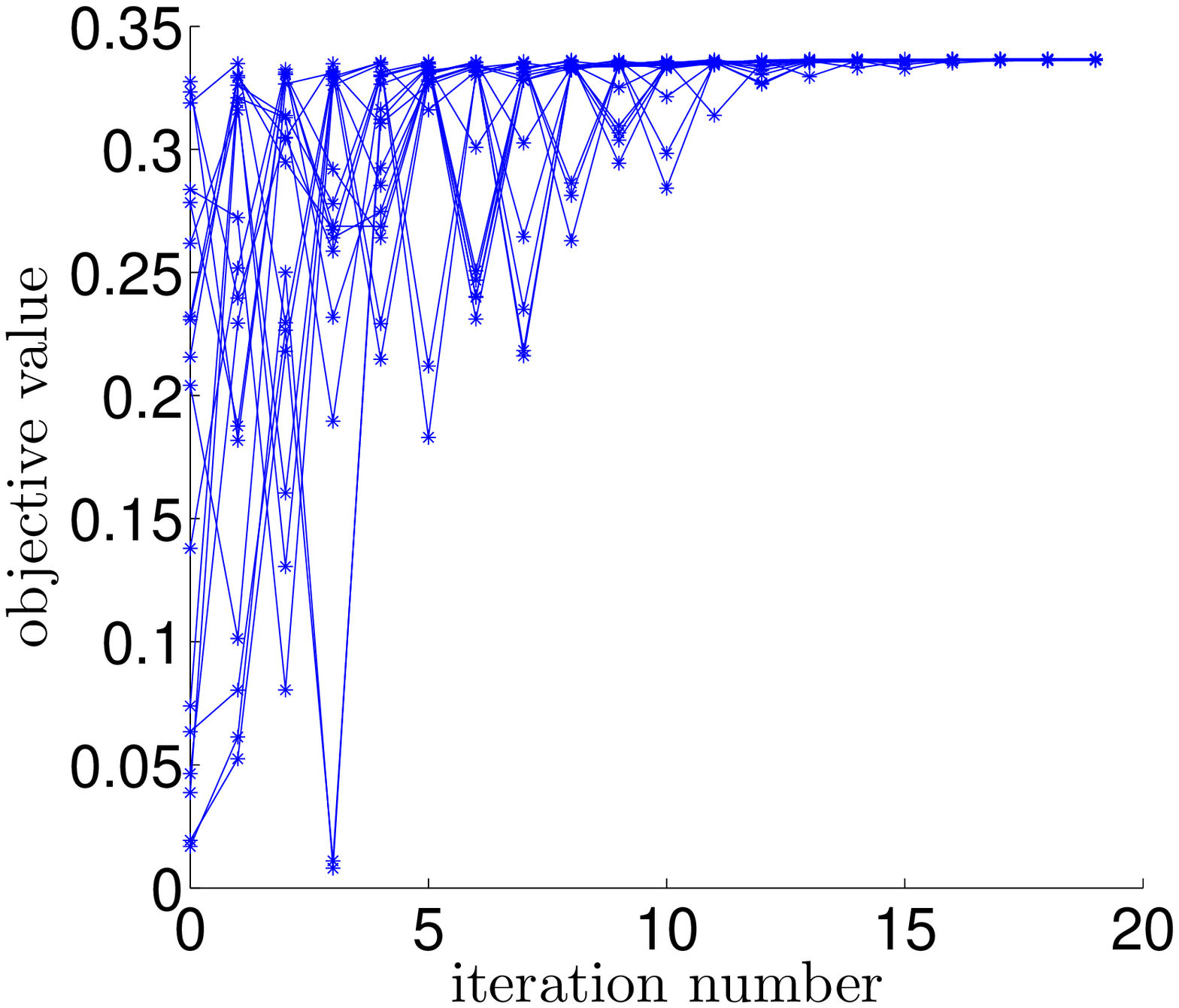}}
\vspace{-8pt}
\caption{Optimization of the objective $\Delta{\bar{\omega}}_{23,\partial B_1, \text{rel}}$ in the absence of resonator: (a) optimal Floquet-Bloch spectrum; (b) convergent objective functions vs iterations for different quasi-Monte Carlo initializations.}
 \label{MODFig02}
\end{figure}

Considering that the first two dispersion curves always meet at the origin for the selected choice of the parameter range (so that the case $h=1$ and $k=2$ has no solution), in the absence of the resonator the optimization problem reads
\begin{eqnarray}\label{eq:problem1}
&\underset{\boldsymbol{\bar{\mu }}^s}{\rm maximize}& \Delta {{\bar{\omega }}_{23,\partial B_1, \text{rel}}}\left( {\boldsymbol{\bar{\mu }}^s} \right) \nonumber \\
&{\rm s. t.} &  \bar{\mu }^s_{l, \rm min} \leq \bar{\mu }^s_l \leq  \bar{\mu }^s_{l, \rm max}, \,\,l=1,...,4\,, \\
&& {\rm and \,the \,constraints\,} (\ref{eq:constraint1}), (\ref{eq:constraint2}), {\rm and\,} (\ref{eq:constraint3})\,. \nonumber 
\end{eqnarray}
To obtain the numerical results, a quasi-random 100-points Sobol' sequence \cite{Niederreiter1992} in the parameter unit hypercube was generated, then all the points of the subsequence satisfying all the constraints were used as initial points for the method of moving asymptotes. Moreover, the partial relative band gap $\Delta {{\bar{\omega }}_{23,\partial B_1, \text{rel}}}$ was approximated by replacing $\partial B_1$ with its uniform discretization, using 30 points. After each valid (constrains-compatible) quasi-Monte Carlo initialization, a number of iterations of the method of moving asymptotes sufficiently large to obtain convergence was performed. The results of the optimization are reported in Fig.\ref{MODFig02}, and demonstrate the presence of a partial relative band gap (with amplitude approximately equal to 0.337) at the {\em best} (higher-valued) objective $\Delta {\bar{\omega }}_{23}^*$. The associated optimal parameter set $\boldsymbol{\bar{\mu }}^*$ is listed in the first row of Table \ref{MODTab02}.

Then, the problem (\ref{eq:problem1}) has been extended to the optimization of the full relative band gap between the second and third dispersion curves. A uniform 90-point discretization of the boundary $\partial B$ has been employed, and zero has been obtained as best value of the objective, corresponding to the absence of full band-gap. This result has been also confirmed by evaluating the objective function on a sufficiently fine grid in the parameter space (10 points for each component), considering only the admissible range of the constrained parameters. 

\begin{figure}
\subfigure[]{\includegraphics[scale=0.33]{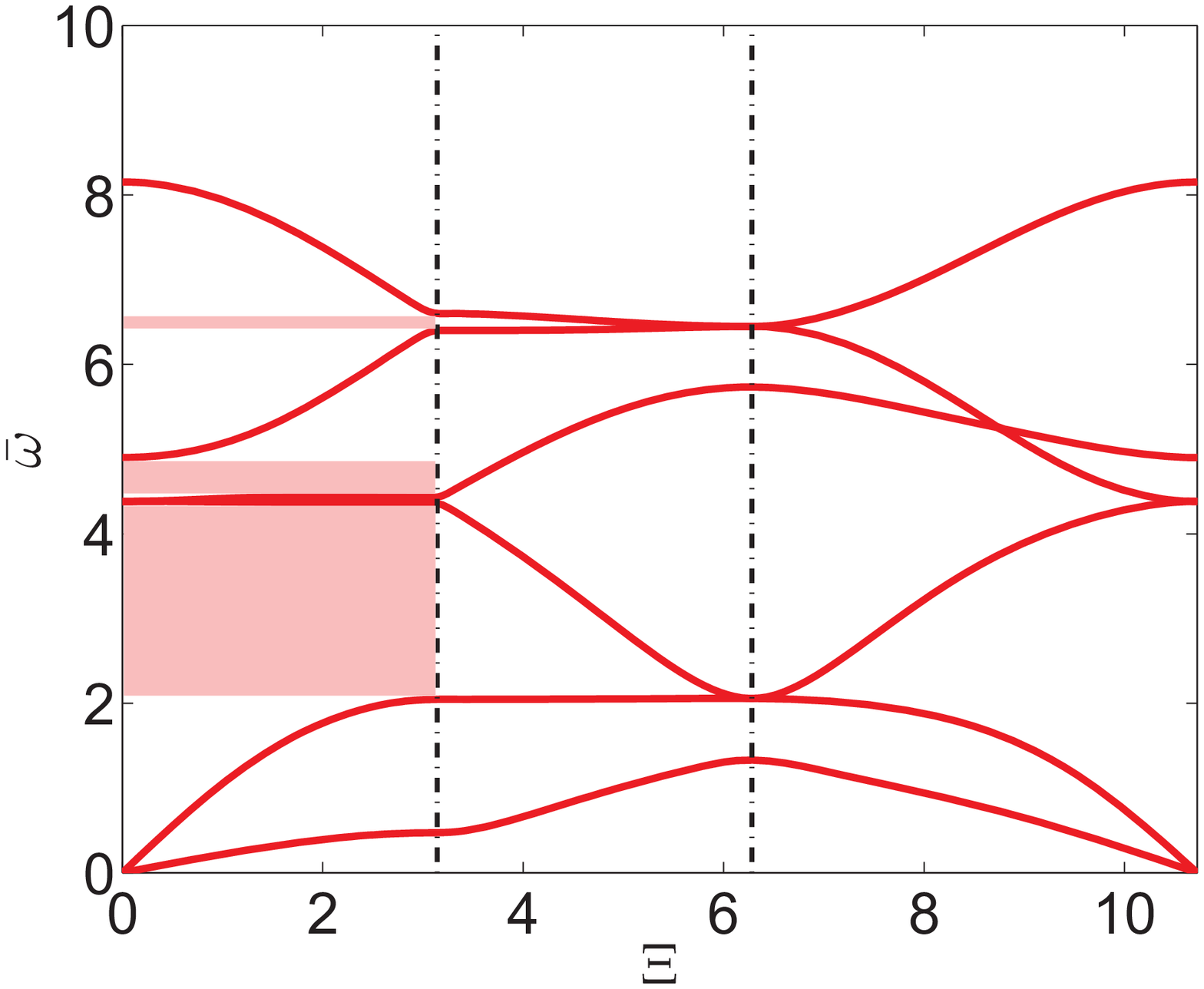}}
\hspace{0.5cm}
\subfigure[]{\includegraphics[scale=0.33]{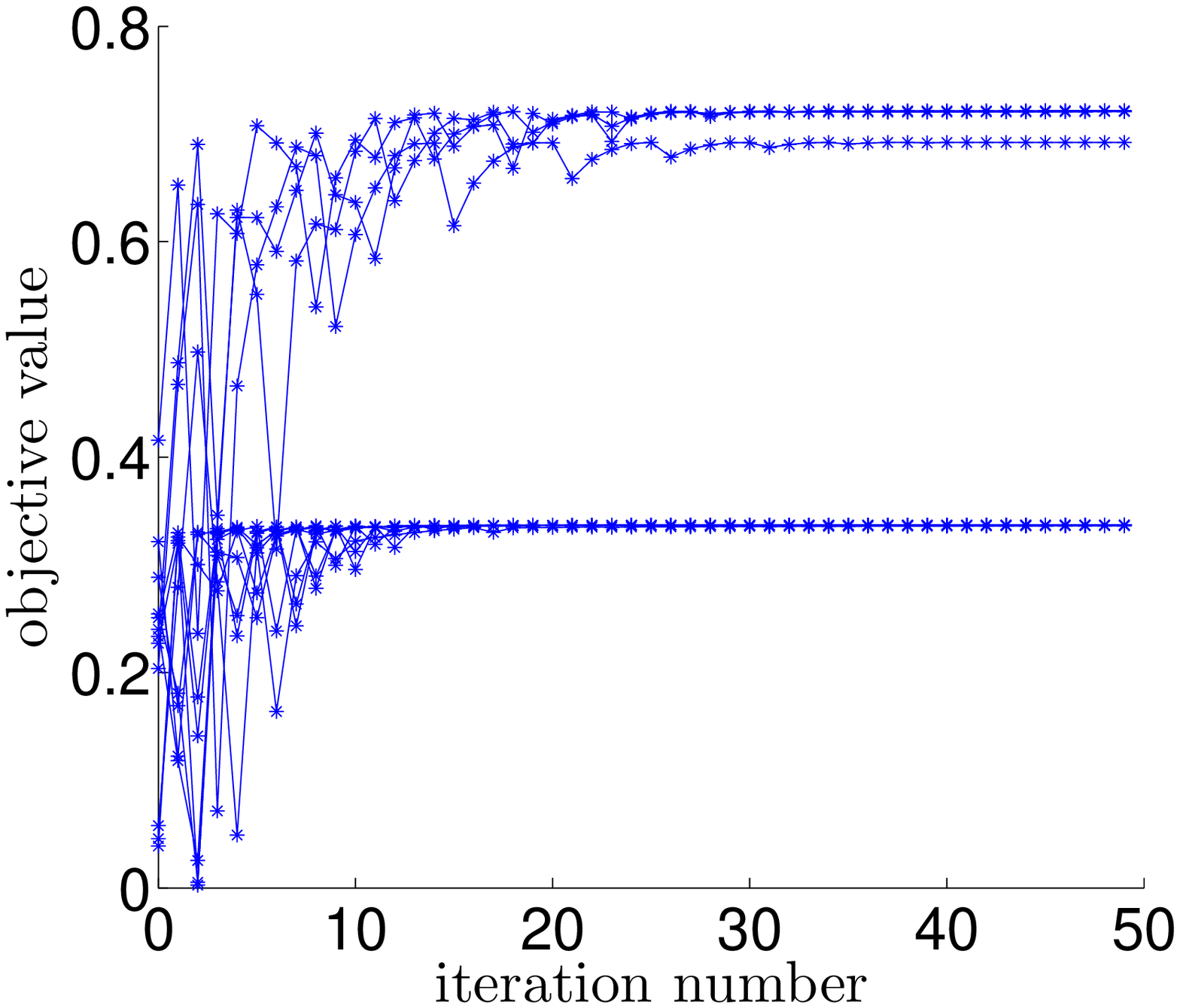}}
\vspace{-8pt}
\caption{Optimization of the objective $\Delta{\bar{\omega}}_{23,\partial B_1, \text{rel}}$ in the presence of resonator: (a) optimal Floquet-Bloch spectrum; (b) convergent objective functions vs iterations for different quasi-Monte Carlo initializations.}
 \label{MODFig03}
\end{figure}

The presence of a partial relative band gap between the second and third dispersion curves has been also obtained as result of the optimization problem 
\begin{eqnarray}\label{eq:problem2}
&\underset{\boldsymbol{\bar{\mu }}}{\rm maximize}& \Delta {{\bar{\omega }}_{23,\partial B_1, \text{rel}}}\left( {\boldsymbol{\bar{\mu }}} \right) \nonumber \\
&{\rm s. t.} &  \bar{\mu }_{l, \rm min} \leq \bar{\mu }_l \leq  \bar{\mu }_{l, \rm max}, \,\,l=1,...,8\,, \\
&& {\rm and \,the \,constraints\,} (\ref{eq:constraint1}), (\ref{eq:constraint2}), (\ref{eq:constraint3}), {\rm and\,} (\ref{eq:constraint4})\,, \nonumber
\end{eqnarray}
which rises up with the introduction of the resonator (see Fig. \ref{MODFig03}). In this case, the method of moving asymptotes has converged to various solutions characterized by different objective values, and the best partial relative band gap has been found approximately equal to $\Delta {\bar{\omega }}_{23}^*=0.722$. This result demonstrates that the presence of the resonator can increase the optimal band gap amplitude. Two more partial relative band gaps, between the fourth-fifth and fifth-sixth pairs of dispersion curves have been obtained (Fig. \ref{MODFig03}a). The associated optimal parameter set $\boldsymbol{\bar{\mu }}^*$ is listed in the second row of Table \ref{MODTab02}.

Finally, the problem (\ref{eq:problem2}) has been extended to the optimization of the {\em full} (with $\partial B$ instead of $\partial B_1$) relative band gap between the second and third dispersion curves. Again, the best value of the objective obtained by the combined method has resulted to be zero, meaning that the presence of the resonator is unable to open a full band-gap between the second and third dispersion curves. 

\subsection{Weighted\,band\,gap\,between\,the\,third\,and\,forth\,dispersion\,curves}

\begin{figure}[t]
\vspace{-5 pt}
\subfigure[]{\includegraphics[scale=0.33]{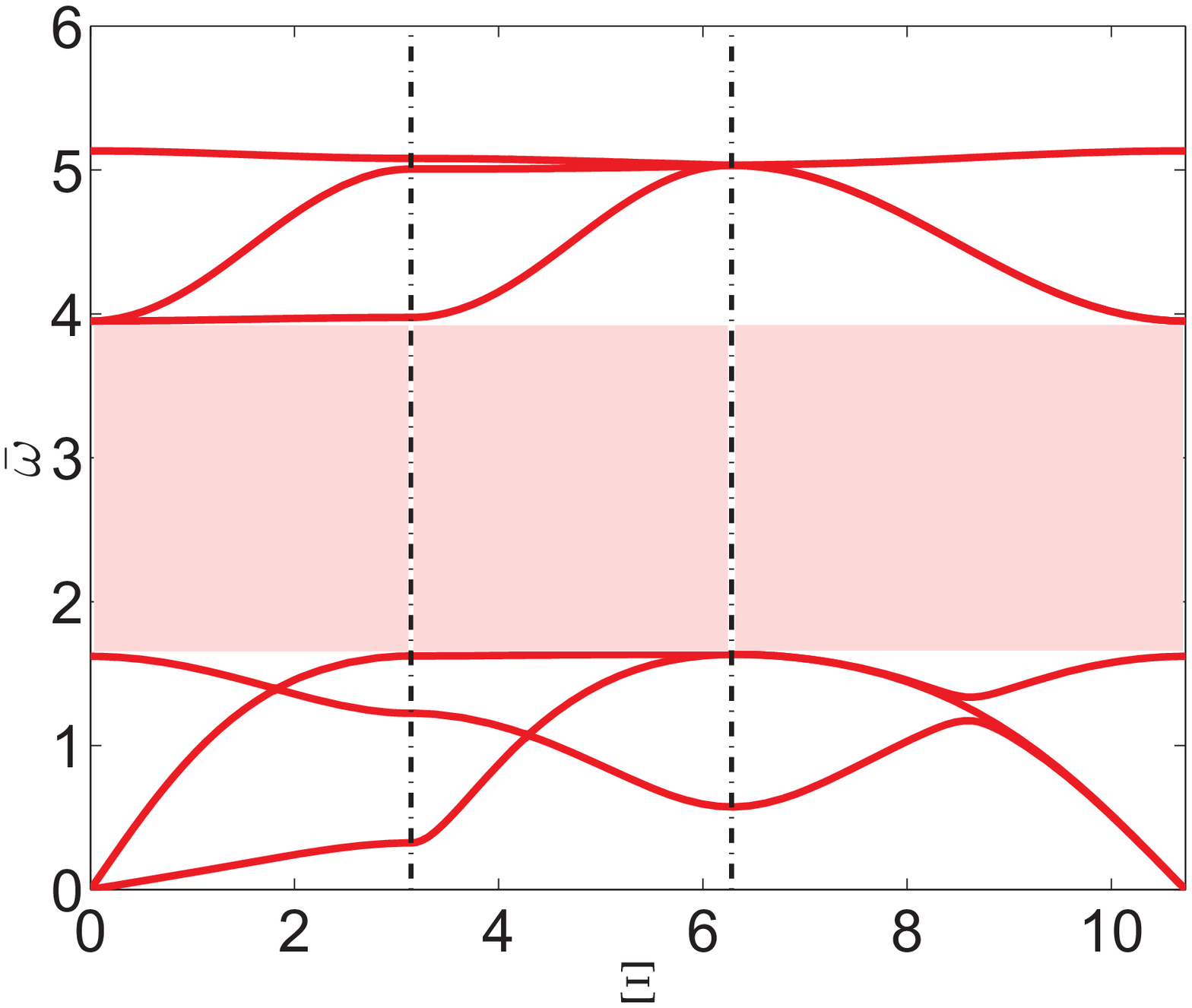}}
\hspace{0.5cm}
\subfigure[]{\includegraphics[scale=0.33]{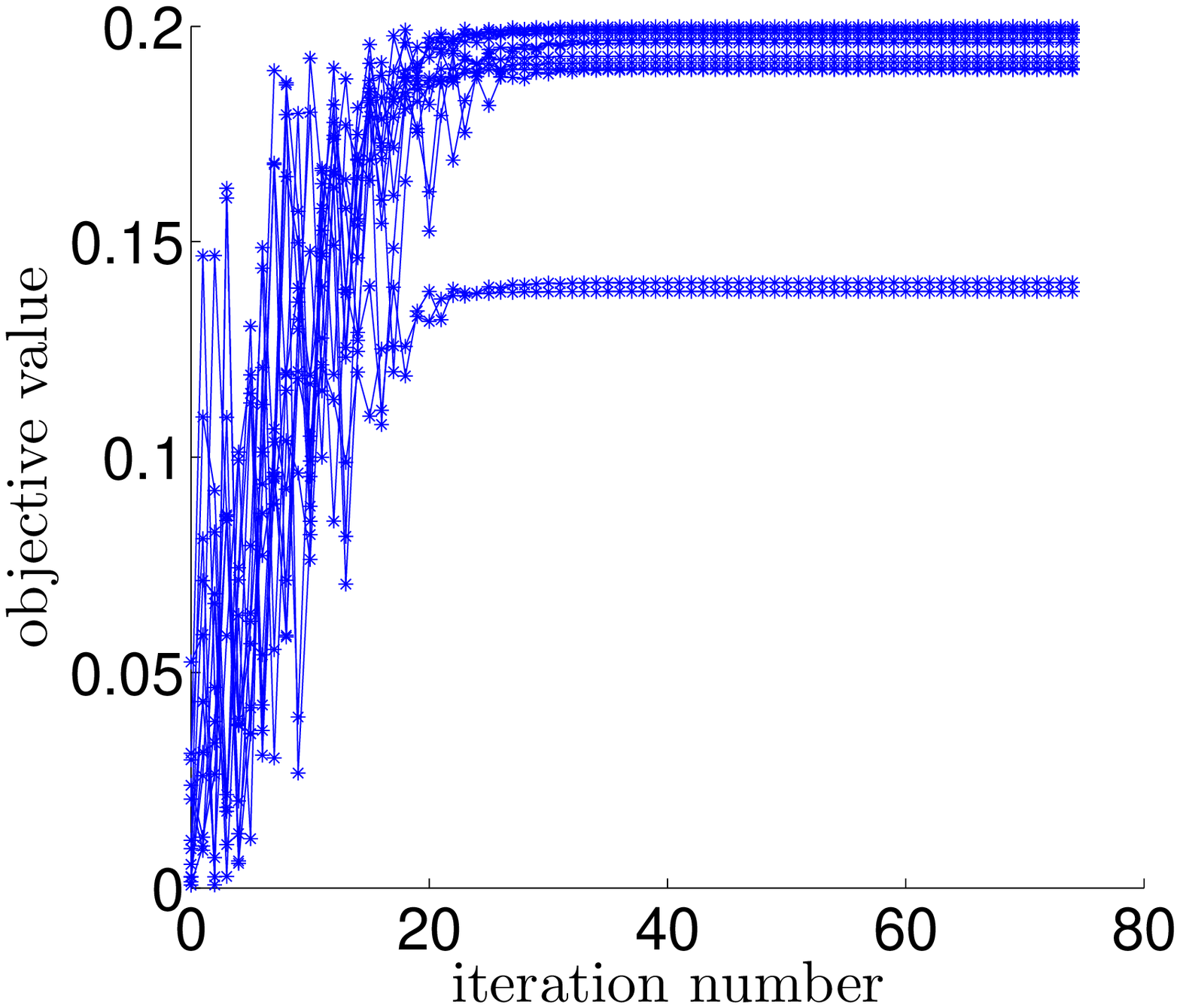}}
\vspace{-5 pt}
\caption{Optimization of the objective $\Delta{\bar{\omega}}_{34,\partial B, \text{rel}}$ in the presence of resonator: (a) optimal Floquet-Bloch spectrum; (b) convergent objective functions vs iterations for different quasi-Monte Carlo initializations.}
 \label{MODFig04}
\end{figure}

The full relative band gap between the third and fourth dispersion curves has been considered in the presence of resonator. To this aim, the optimization problem is re-formulated to maximize a trade-off (i.e., the product\footnote{In a preliminary phase, we also considered as objective function a weighted sum, with positive weights, of the full relative band gap and the fluctuation of the fourth dispersion curve. However, for various choices of the weights, the obtained solution was characterized either by a negligible value of the full relative band gap, or by a negligible value of the fluctuation of the fourth dispersion curve, making an optimal choice of the weights difficult. For our specific goal, instead, the product of the two terms was more effective.}) between the full relative band gap, and the {\em band amplitude} $\Delta_A{\bar{\omega }}_{4,\partial B}=\max_{\partial B}({\bar{\omega}}_{4})-\min_{\partial B}({\bar{\omega}}_{4})$ of the fourth dispersion curve
\begin{eqnarray}\label{eq:problem3}
&\underset{\boldsymbol{\bar{\mu }}}{\rm maximize}& \left(\Delta {{\bar{\omega }}_{34,\partial B, \text{rel}}}\left( {\boldsymbol{\bar{\mu }}} \right)\;\Delta_A {{\bar{\omega }}_{4,\partial B, \text{rel}}}\left( {\boldsymbol{\bar{\mu }}} \right) \right) \nonumber \\
&{\rm s. t.} &  \bar{\mu }_{l, \rm min} \leq \bar{\mu }_l \leq  \bar{\mu }_{l, \rm max}, \,\,l=1,...,8\,, \\
&& {\rm and \,the \,constraints\,} (\ref{eq:constraint1}), (\ref{eq:constraint2}), (\ref{eq:constraint3}), {\rm and\,} (\ref{eq:constraint4})\,. \nonumber
\end{eqnarray}
The reason is that, in the absence of an elastic coupling between the resonator and the cell structure, three of the dispersion curves are $\mathrm\Xi$-independent (null band amplitude), as they express the fixed frequencies of the free-standing resonator. Therefore, a large but definitely not significant relative band gap would be obtained by separating such curves from the rest of the Floquet-Bloch spectrum as much as possible. The relative band amplitude of the fourth dispersion curve, acting as a weighting multiplier in the objective function, is expected to avoid this shortcoming. Indeed, by taking the product, preference is given to parametric designs of beam-lattice metamaterials for which both factors $\Delta {{\bar{\omega }}_{34,\partial B, \text{rel}}}\left( {\boldsymbol{\bar{\mu }}} \right)$ and $\Delta_A {{\bar{\omega }}_{4,\partial B, \text{rel}}}\left( {\boldsymbol{\bar{\mu }}} \right)$ are simultaneously large. The optimization results are reported in Fig.\ref{MODFig04}. The best solution is associated with a full band gap with relative amplitude approximately equal to $0.830$, whereas the relative band amplitude of the fourth dispersion curve is about $0.241$. The optimal values of the parameters are listed in the third row of Table \ref{MODTab02}.

\begin{table}[b]
\vspace{0pt}
\begin{center}
\setlength{\tabcolsep}{8pt}
\renewcommand{\arraystretch}{1.5}
\begin{tabular}{c c c c c c c c c}
\hline
\rowcolor{Gray}
{\rule{-1pt}{12pt}\rm Fig.} & $\frac{w}{\varepsilon}$ & $\frac{w_{an}}{w}$ & $\frac{R}{\varepsilon}$ & $\beta$ & $\frac{r}{\varepsilon}$ & $\frac{E_r}{E_s}$ & $\nu_r$ & $\frac{\rho_r}{\rho_{s\_an}}\rule[-8pt]{-1pt}{12pt}$ \\
\hline
$\rule{-1pt}{12pt} \ref{MODFig02}$ & $0.0600$ & $3.19$ & $0.200$ & $0.295$ & $-$ & $-$ & $-$ & $-\rule[-8pt]{-1pt}{12pt}$ \\
\hline
$\rule{-1pt}{12pt} \ref{MODFig03}$ & $0.139$ & $0.143$ & $0.198$ & $0.404$ & $0.0864$ & $0.100$ & $0.200$ & $2.00\rule[-8pt]{-1pt}{12pt}$ \\
\hline
$\rule{-1pt}{12pt} \ref{MODFig04}$ & $0.0950$ & $0.439$ & $0.200$ & $0.00163$ & $0.127$ & $0.104$ & $0.218$ & $1.93\rule[-8pt]{-1pt}{12pt}$ \\
\hline
\end{tabular}
\end{center}
\caption{Best parameters obtained in the optimization reported in Fig. \ref{MODFig02}, \ref{MODFig03}, and \ref{MODFig04}.}
\label{MODTab02}
\end{table}

Summing up, the main results show that, for a periodic cell with fixed characteristic size, the presence of resonators is a mandatory condition for the existence of full band gaps in the low-frequency range. However, only partial band-gap can be opened between the second and third branches. Full band-gaps can be obtained between the third and forth branches, and -- by virtue of the optimization -- the largest amplitude can overcome the maximum frequency of the acoustic or first optical branches of the spectrum. The optimal results correspond large-radius rings and highly-slender, non-tangent ligaments with quasi-negligible inclination (corresponding to a nearly-vanishing geometric chirality). Accordingly, the optimized resonators are found to possess about half the radius of the rings and be embedded in a highly-soft matrix.

\section{Conclusions}\label{sec:4}
A parametric model of periodic metamaterial has been formulated, and its wave-propagation properties have been investigated. Then, some optimization problems related to such properties have been stated and solved numerically. From the physical viewpoint, the desirable target is a high-performing material with marked filtering capacities for low-frequency signals. The optimal results demonstrate that a partial band gap can be obtained between the second and third dispersion curves, both in the absence and in the presence of a resonator inside the periodic cell. However, in both cases, no full relative band gaps are obtainable. On the contrary, the resonator allows the opening of a positive full relative band gap between the third and fourth dispersion curves, associated with a non-negligible elastic coupling between the resonator and the cell structure. This achievement implied the optimization of a multiplicative trade-off between the band gap and the relative band amplitude of the fourth dispersion curve. All the optimization problems have been solved numerically by combining the method of moving asymptotes with a quasi-Monte Carlo multi-start technique. 

As possible developments, the optimization framework could be extended to its regularized version, to reduce the sensitivity of the obtained solution with respect to changes in the nominal parameter values. To this aim, suitable regularization techniques, typical of machine learning problems, could be used \cite{GneccoSanguineti2010,GneccoSanguineti2013}. Other nonlinear optimization methods could be also used, such as sequential linear or quadratic programming. Finally, an electromechanical extension of the physical-mathematical model would allow the design of smart metamaterials.

\section*{Appendix}\label{sec:appendix}

Adapting to the present tetrachiral context the analysis made in \cite{Bacigalupoetal2016} for the hexachiral case, one can show that the non vanishing components of the three-by-three positive definite diagonal submatrices ${\mathbf{\bar{M}}}^{s}$ and ${\mathbf{\bar{M}}}^{r}$, that make up the six-by-six dimensionless block diagonal matrix ${\mathbf{\bar{M}}}$, 
read
\begin{gather}
\bar{M}^s_{11}= 2\,\pi \,\frac{R}{\varepsilon}\,\frac{w}{\varepsilon}\,\frac{w_{an}}{w},\qquad
\bar{M}^s_{22}= 2\,\pi \,\frac{R}{\varepsilon}\,\frac{w}{\varepsilon}\,\frac{w_{an}}{w},\\
\bar{M}^s_{33}= \frac{1}{2}\,\pi \,\frac{R}{\varepsilon}\,\frac{w}{\varepsilon}\,\frac{w_a}{w}\, \Big( \left(\frac{w}{\varepsilon}\right)^{2}\left(\frac{w_a}{w}\right)^{2}+4\,\Big(\frac{R}{\varepsilon}\Big)^{2} \Big),\nonumber\\
\bar{M}^r_{11}= \pi \,\left({\frac{r}{\varepsilon}}\right)^{2}\frac{\rho_r}{\rho_{s\_an}},\quad
\bar{M}^r_{22}= \pi \,\left({\frac{r}{\varepsilon}}\right)^{2}\frac{\rho_r}{\rho_{s\_an}},\quad
\bar{M}^r_{33}= \frac{1}{2}\,\pi \,\frac{\rho_r}{\rho_{s\_an}}\,\left({\frac{r}{\varepsilon}}\right)^{4}.\nonumber
\end{gather}
In order to introduce the components of the three-by-three Hermitian submatrix ${\mathbf{\bar{K}}}^{s}$, we need to introduce the dependent parameters ${k_d}/{E_s}$ and ${k_\theta}/({\varepsilon^2 E_s})$, which are functions of the other parameters, respectively, of the form $\frac{k_d}{E_s}=f_d\left(\frac{r}{\varepsilon} \frac{\varepsilon}{R}, \frac{E_r}{E_s}, \nu_r\right)$ and $\frac{k_\theta}{\varepsilon^2 E_s}=f_\theta \left(\frac{r}{\varepsilon} \frac{\varepsilon}{R}, \frac{E_r}{E_s}, \nu_r\right)$. The definitions of these functions are reported in \cite{Bacigalupoetal2016}. Then, the components of ${\mathbf{\bar{K}}}^{s}$ are expressed as follows:
\begin{gather}
\bar{K}_{ij}^{s}=\frac{1}{\Lambda }\left( \bar{K}_{ij}^{s\_3}{{\left( \frac{w}{\varepsilon} \right)}^{3}}+ \bar{K}_{ij}^{s\_1}\frac{w}{\varepsilon}+ \bar{K}_{ij}^{s\_0} \right)\,,\end{gather}
where $i,j=1,2,3$, $\Lambda=\left( \cos \left( \beta \right) -\Psi \right) ^{3}$ and $\Psi=\sqrt { \left( \cos \left( \beta \right)  \right) ^{2}+4\,{\left(\frac{R}{\varepsilon}\right)}^{2}-1}$. More precisely, one obtains
\begin{align}
\bar{K}_{11}^{s\_3}&=\left( 2\,\cos \left( \bar{k}_1 \right) -2\,\cos \left( \bar{k}_2 \right)  \right)  \left( \cos \left( \beta \right)  \right) ^{
2}-2\,\cos \left( \bar{k}_1 \right) +2\,\\
\bar{K}_{11}^{s\_1} &= \left( -2\,\cos \left( \bar{k}_1 \right) +2\,\cos \left(  \bar{k}_2 \right)  \right)  \left( \cos \left( \beta \right)  \right) ^{
4} \nonumber \\
& + \left( 4\,\cos \left( \bar{k}_1 \right) \Psi-4\,\cos \left(  \bar{k}_2 \right) \Psi \right)  \left( \cos \left( \beta \right) 
 \right) ^{3} \nonumber \\
& + \left( -2\,\cos \left( \bar{k}_1 \right) {\Psi}^{2}+2
\,\cos \left( \bar{k}_2 \right) {\Psi}^{2}-2\,\cos \left( \bar{k}_2 \right) +2 \right)  \left( \cos \left( \beta \right)  \right) 
^{2} \nonumber \\
& + \left( 4\,\cos \left( \bar{k}_2 \right) \Psi-4\,\Psi \right) 
\cos \left( \beta \right) -2\,\cos \left( \bar{k}_2 \right) {\Psi}^{
2}+2\,{\Psi}^{2} \,, \nonumber\\
\bar{K}_{11}^{s\_0}&= \left( \cos \left( \beta \right)  \right) ^{3}{\frac{k_d}{E_s}}-3\,
 \left( \cos \left( \beta \right)  \right) ^{2}\Psi\,{\frac{k_d}{E_s}}+3\,
\cos \left( \beta \right) {\Psi}^{2}{\frac{k_d}{E_s}}-{\Psi}^{3}{\frac{k_d}{E_s}}\,,\nonumber\\
\bar{K}_{22}^{s\_3}&=\left( -2\,\cos \left( \bar{k}_1 \right) +2\,\cos \left( \bar{k}_2 \right)  \right)  \left( \cos \left( \beta \right)  \right) ^{
2}-2\,\cos \left( \bar{k}_2 \right) +2\,,\nonumber\\
\bar{K}_{22}^{s\_1} &=\left( 2\,\cos \left( \bar{k}_1 \right) -2\,\cos \left( \bar{k}_2 \right)  \right)  \left( \cos \left( \beta \right)  \right) ^{
4} \nonumber \\ 
& + \left( -4\,\cos \left( \bar{k}_1 \right) \Psi+4\,\cos \left( \bar{k}_2 \right) \Psi \right)  \left( \cos \left( \beta \right) 
 \right) ^{3} \nonumber \\
& + \left( 2\,\cos \left( \bar{k}_1 \right) {\Psi}^{2}-2
\,\cos \left( \bar{k}_2 \right) {\Psi}^{2}-2\,\cos \left( \bar{k}_1 \right) +2 \right)  \left( \cos \left( \beta \right)  \right) 
^{2} \nonumber \\
& + \left( 4\,\cos \left( \bar{k}_1 \right) \Psi-4\,\Psi \right) 
\cos \left( \beta \right) -2\,\cos \left( \bar{k}_1 \right) {\Psi}^{
2}+2\,{\Psi}^{2}\,, \nonumber\\
\bar{K}_{22}^{s\_0}&=\left( \cos \left( \beta \right)  \right) ^{3}\frac{k_d}{E_s}-3\,
 \left( \cos \left( \beta \right)  \right) ^{2}\Psi\,\frac{k_d}{E_s}+3\,
\cos \left( \beta \right) {\Psi}^{2}\frac{k_d}{E_s}-{\Psi}^{3}\frac{k_d}{E_s}\nonumber\\
\bar{K}_{33}^{s\_3}&=\left( 1/3\,\cos \left( \bar{k}_1 \right) +1/3\,\cos \left( \bar{k}_2 \right) +4/3 \right)  \left( \cos \left( \beta \right) 
 \right) ^{2} \nonumber \\
& + \left( -2/3\,\Psi+1/3\,\cos \left( \bar{k}_1 \right) 
\Psi+1/3\,\cos \left( \bar{k}_2 \right) \Psi \right) \cos \left( 
\beta \right) \nonumber \\
& -1/6\,\cos \left( \bar{k}_1 \right) {\Psi}^{2} -1/6\,
\cos \left( \bar{k}_2 \right) {\Psi}^{2}+1/3\,{\Psi}^{2}\,, \nonumber\\
\bar{K}_{33}^{s\_1}&= \left( -1/2\,\cos \left( \bar{k}_1 \right) -1/2\,\cos \left( \bar{k}_2  \right) -1 \right)  \left( \cos \left( \beta \right)  \right) 
^{4} \nonumber \\
& + \left( \cos \left( \bar{k}_1 \right) \Psi+\cos \left( \bar{k}_2 \right) \Psi+2\,\Psi \right)  \left( \cos \left( \beta \right) 
 \right) ^{3} \nonumber \\
 & + \bigg( 1/2\,\cos \left( \bar{k}_2 \right) +1/2\,\cos
 \left( \bar{k}_1 \right) +1-{\Psi}^{2}-1/2\,\cos \left( \bar{k}_1 \right) {\Psi}^{2} \nonumber \\
& \,\,\,\,\,\,\,\,\,\,\,\, -1/2\,\cos \left( \bar{k}_2 \right) {\Psi}^{2}
 \bigg)  \left( \cos \left( \beta \right)  \right) ^{2} \nonumber \\
 & + \left( -\cos
 \left( \bar{k}_1 \right) \Psi-\cos \left( \bar{k}_2 \right) \Psi-
2\,\Psi \right) \cos \left( \beta \right) \nonumber \\
& +1/2\,\cos \left( \bar{k}_1 \right) {\Psi}^{2}+1/2\,\cos \left( \bar{k}_2 \right) {\Psi}^{2}+{
\Psi}^{2}\,, \nonumber\\
\bar{K}_{33}^{s\_0}&=-3\, \left( \cos \left( \beta \right)  \right) ^{2}\Psi\,\frac{k_\theta}{\varepsilon^2 E_s}+3\,\cos \left( \beta \right) {\Psi}^{2}\frac{k_\theta}{\varepsilon^2 E_s}+
 \left( \cos \left( \beta \right)  \right) ^{3}\frac{k_\theta}{\varepsilon^2 E_s}-{\Psi}^
{3}\frac{k_\theta}{\varepsilon^2 E_s}\,,\nonumber\\
\bar{K}_{12}^{s\_3}&=2\,\cos \left( \beta \right) \sin \left( \beta \right)  \left( -\cos
 \left( \bar{k}_1 \right) +\cos \left( \bar{k}_2 \right) 
 \right)\,,\nonumber\\
\bar{K}_{12}^{s\_1}&=2\,\sin \left( \beta \right)  \left( \cos \left( \bar{k}_1 \right) 
-\cos \left( \bar{k}_2 \right)  \right)  \left( \cos \left( \beta
 \right)  \right) ^{3} \nonumber \\
 & +2\,\sin \left( \beta \right)  \left( -2\,\cos
 \left( \bar{k}_1 \right) \Psi+2\,\cos \left( \bar{k}_2 \right) 
\Psi \right)  \left( \cos \left( \beta \right)  \right) ^{2} \nonumber \\
& +2\,\sin
 \left( \beta \right)  \left( \cos \left( \bar{k}_1 \right) {\Psi}^{
2}-\cos \left( \bar{k}_2 \right) {\Psi}^{2} \right) \cos \left( 
\beta \right)
\,, \nonumber\\
\bar{K}_{12}^{s\_1}&=0\,,\nonumber\\
\bar{K}_{13}^{s\_3}&=-i\sin \left( \bar{k}_2 \right)  \left( \cos \left( \beta \right) 
 \right) ^{2}+i\sin \left( \beta \right) \sin \left( \bar{k}_1
 \right) \cos \left( \beta \right)\,,\nonumber\\
\bar{K}_{13}^{s\_1}&=-i\sin \left( \bar{k}_2 \right)  \left( \cos \left( \beta \right) 
 \right) ^{4}-i \left( -\sin \left( \beta \right) \sin \left( \bar{k}_1 \right) -2\,\sin \left( \bar{k}_2 \right) \Psi \right) 
 \left( \cos \left( \beta \right)  \right) ^{3} \nonumber \\
 & -i \left( 2\,\sin
 \left( \bar{k}_1 \right) \sin \left( \beta \right) \Psi+\sin
 \left( \bar{k}_2 \right) {\Psi}^{2}-\sin \left( \bar{k}_2
 \right)  \right)  \left( \cos \left( \beta \right)  \right) ^{2} \nonumber \\
 & -i
 \left( -\sin \left( \bar{k}_1 \right) \sin \left( \beta \right) {
\Psi}^{2}+2\,\sin \left( \bar{k}_2 \right) \Psi \right) \cos \left( 
\beta \right) +i\sin \left( \bar{k}_2 \right) {\Psi}^{2}
\,, \nonumber
\end{align}
\begin{align}
\bar{K}_{13}^{s\_1}&=0\,,\nonumber\\
\bar{K}_{23}^{s\_3}&=i\sin \left( \bar{k}_1 \right)  \left( \cos \left( \beta \right) 
 \right) ^{2}+i\sin \left( \beta \right) \sin \left( \bar{k}_2
 \right) \cos \left( \beta \right)\,,\nonumber\\
\bar{K}_{23}^{s\_1}&=i\sin \left( \bar{k}_1 \right)  \left( \cos \left( \beta \right) 
 \right) ^{4}+i \left( \sin \left( \beta \right) \sin \left( \bar{k}_2 \right) -2\,\sin \left( \bar{k}_1 \right) \Psi \right) 
 \left( \cos \left( \beta \right)  \right) ^{3} \nonumber \\
& +i \left( -2\,\sin
 \left( \beta \right) \sin \left( \bar{k}_2 \right) \Psi+\sin
 \left( \bar{k}_1 \right) {\Psi}^{2}-\sin \left( \bar{k}_1
 \right)  \right)  \left( \cos \left( \beta \right)  \right) ^{2} \nonumber \\
 & +i
 \left( \sin \left( \beta \right) \sin \left( \bar{k}_2 \right) {
\Psi}^{2}+2\,\sin \left( \bar{k}_1 \right) \Psi \right) \cos \left( 
\beta \right) -i\sin \left( \bar{k}_1 \right) {\Psi}^{2}
\,, \nonumber\\
 \bar{K}_{23}^{s\_1}&=0\,,\nonumber \\
 \bar{K}_{21}^{s}&=\bar{K}_{12}^{s}\,\nonumber \\
 \bar{K}_{31}^{s}&=-i {\rm Im}\left(\bar{K}_{13}^{s}\right)\,,\nonumber \\
 \bar{K}_{32}^{s}&=-i {\rm Im}\left(\bar{K}_{23}^{s}\right)\,,\nonumber
\end{align}
where $i$ denotes the imaginary unit and ${\rm Im}(z)$ denotes the imaginary part of the complex number $z$.
Finally, the non vanishing components of the diagonal submatrix ${\mathbf{\bar{K}}}^{r}$ are
\begin{align}
\bar{K}_{11}^{r}= \frac{k_d}{E_s}\,,&& \bar{K}_{22}^{r}= \frac{k_d}{E_s}\,,&& \bar{K}_{33}^{r}= \frac{k_\theta}{\varepsilon^2 E_s}
\end{align}
whereas the diagonal submatrices ${\mathbf{\bar{K}}}^{sr}$ and ${\mathbf{\bar{K}}}^{rs}$ satisfy the constraint ${\mathbf{\bar{K}}}^{sr}=\left({\mathbf{\bar{K}}}^{sr}\right)^T=-{\mathbf{\bar{K}}}^{r}$.

\end{document}